\documentstyle[aps,pra,twocolumn]{revtex}
\begin{document}
\draft

\newcommand{\pp}[1]{\phantom{#1}}
\newcommand{\be}{\begin{eqnarray}}
\newcommand{\ee}{\end{eqnarray}}
\newcommand{\ve}{\varepsilon}
\newcommand{\vs}{\varsigma}
\newcommand{\Tr}{{\,\rm Tr\,}}
\newcommand{\pol}{\frac{1}{2}}

\title{
Local modification of the Abrams-Lloyd nonlinear algorithm
}
\author{Marek Czachor}
\address{
Katedra Fizyki Teoretycznej i Metod Matematycznych\\
 Politechnika Gda\'{n}ska,
ul. Narutowicza 11/12, 80-952 Gda\'{n}sk, Poland
}
\maketitle

\begin{abstract}
The nonlinear algorithms proposed recently by Abrams and Lloyd
[Report No. quant-ph/9801041] are fast but make an explicit use
of an arbitrarily fast unphysical transfer of information within
a quantum computer. It is shown that there exists a
simplification of the second Abrams-Lloyd algorithm which
eliminates the unphysical effect but keeps the algorithm fast. 
\end{abstract}

\narrowtext
\section{Introduction}

Any systematic procedure associating with an arbitrary number $i_0\dots
i_{n-1}$, $i_k=0$ or 
1, another number $f(i_0\dots i_{n-1})$ can be termed an algorithm.
The idea of quantum computation rests on the observation
that a binary number $i_0\dots i_{n-1}$
can be represented by a vector (a qubinary number) $|i_0\rangle\dots
|i_{n-1}\rangle$ representing an uncorrelated state of $n$
distinguishable two-level quantum systems. 
A quantum algorithm is essentially an algorithm based on a qubinary
representation of numbers \cite{Deutsch,Feynman}. 

Typical quantum algorithms use unitary operations and
projections \cite{Bennett,FT}. 
This is motivated by the unitarity of the standard
Schr\"odinger dynamics and the so-called projection postulate. 
The latter postulate is typical of the Copenhagen interpretation
of quantum mechanics and is not essential to quantum
computation (quantum algorithms would look different
in, say, the Many Worlds interpretation of quantum mechanics but
their fundamental properties would not change). 
The Schr\"odinger (linear and unitary) dynamics is not the only
dynamics one encounters in quantum theories. Dynamics in
the Heisenberg picture is typically nonlinear (in the sense that
operators depend nonlinearly on initial conditions). Typical
effective dynamics of quantum optical systems (such as two-level
atoms) is irreversible and nonunitary. There are also
many situations where states of quantum systems evolve in an
effectively nonlinear way (optical solitons, Hartree-type
approximations). Finally, there is still no proof that the present-day
quantum mechanics is not an approximation to a more exact
nonlinear theory, and various versions of such a nonlinear
generalization have been proposed. It should be stressed that
the popular oppinion stating that all nonlinear extensions of quantum
mechanics must lead to logical absurdities does not find support
in a detailed analysis of nonlinear ``no-go" theorems (for a
brief discussion cf. \cite{MCrc}).

The 
assumption of fundamental quantum linearity cannot be regarded
as a consquence of experimental 
data because it is quite typical that a consistent theory is
prior to experiments. This point was clear to Wigner
\cite{Wigner39} who was simultaneously one of the first to
associate fundamentally nonlinear phenomena with a
theory of brain functioning \cite{Wigner-mb}. Since there is no doubt
that human brain is a physical system, there is almost no doubt
that at least some of its aspects have to be described by
quantum mechanics. The idea of qubinary mathematics
(including diferentiation and integration) can be traced back to
Orlov's works 
on a ``wave logic" of conciousness \cite{Orlov}.
On the other hand, there are serious arguments of Penrose 
\cite{Penrose1,Penrose2} for non-algorithmic (in the classical
sense) ingredients in brain activity. Human and animal brains are systems 
that seem to possess a feedback-type property of
self-observation, and feedback effects are typically associated
with a nonlinear evolution. 

Although the above problems may appear somewhat far from standard
quantum physics, they naturally lead to the question of
possible consequences of a nonlinear quantum dynamics for the
theory of quantum computation. 
The problem was recently addressed by Abrams and Lloyd
\cite{al} who showed that a nonlinear evolution in a Hilbert
space of states of a quantum computer leads naturally to
polynomial-time solutions of NP and \#P problems. 
The Abrams-Lloyd argument was based on a general property of
nonlinear evolutions in Hilbert spaces, namely the
non-conservation of scalar products between nonlinearly evolving
solutions of a nonlinear Schr\"odinger equation. This effect
(in the literature called a mobility phenomenon) 
was discussed in great detail by Mielnik
\cite{Mielnik1} in his analysis of
nonlinear motion semigroups (compare also \cite{MP}). 

The mobility effect was used in \cite{al} in two algorithms. In
the first of them the Authors chose a nonlinear evolution
which has $|0\rangle$ 
as a fixed point, but any superposition of
$|0\rangle$ and $|1\rangle$ 
transports towards $|1\rangle$.
In the second algorithm a sequence of nonlinear operations
was partly disentangling
the state of the quantum computer by transformations of the type
\be
\frac{1}{\sqrt{2}}\Big(
|0\rangle|0\rangle+|1\rangle|1\rangle\Big)
\to
\frac{1}{\sqrt{2}}\Big(
|0\rangle|1\rangle+|1\rangle|1\rangle\Big).\label{mob}
\ee
Nonlinear evolutions that lead (with an arbitrary accuracy) to the required
modification of entangled states can be obtained in a
Weinberg-type nonlinear quantum mechanics \cite{Weinberg}.  
There is a problem, however. Using exactly the same trick
[i.e. transformation (\ref{mob})] it was shown in \cite{MCfpl}
that (\ref{mob}) is responsible for arbitrarily fast influences
between noninteracting systems (this
should not be confused with another effect discussed by Gisin
\cite{Gisin}, which was a result of the projection postulate). 
Therefore, the algorithm of Abrams and Lloyd makes an explicit
use of an unphysical and arbitrarily fast process, so there is a
danger that this is the reason why it is fast. 

Fortunately, as we shall see below, this is not the case. 
To prove it we shall concentrate on the second algorithm. We
will consider a concrete example of a nonlinear dynamics and
will use a formalism that is known to eliminate the
unphysical influences \cite{Polchinski,Bona,Jordan,MCpla,MCqph}.
A detailed discussion of both algorithms in this context can be
found in \cite{MCqph2}. 

\section{Second Abrams-Lloyd algorithm}
{\it Step 1.\/} 
We begin with the state 
\be
|\psi[0]\rangle &=&|0_1,\dots,0_n\rangle|0\rangle
\ee
where the first $n$ qubits correspond to the input and the last
qubit represents the output. 

Consider the unitary transformation acting as follows
\be
U|0\rangle &=&\frac{1}{\sqrt{2}}\Big(|0\rangle+ |1\rangle\Big)\\
U|1\rangle &=&\frac{1}{\sqrt{2}}\Big(-|0\rangle+ |1\rangle\Big)
\ee
{\it Step 2.\/} 
\be
|\psi[1]\rangle &=&\underbrace{U\otimes\dots\otimes U}_n\otimes 1
|\psi[0]\rangle\\
&=&\frac{1}{\sqrt{2^n}}\sum_{i_1\dots i_n=0}^1
|i_1,\dots,i_n\rangle|0\rangle
\ee
The input constists now of a uniform superposition of all the
numbers $0\leq n \leq 2^n-1$.

\medskip
\noindent
{\it Step 3.\/} 
\be
|\psi[2]\rangle &=& F|\psi[1]\rangle\\
&=&
\frac{1}{\sqrt{2^n}}\sum_{i_1\dots i_n=0}^1
|i_1,\dots,i_n\rangle|f(i_1,\dots,i_n)\rangle\label{state2}
\ee
where $F$ is some unitary transformation (an oracle) that transforms
the input into an output; $f(i_1,\dots,i_n)$ equals 1 or 0.

\medskip
\noindent
{\it Step 4.\/} 
We assume that $f(x)=1$ for at most one $x$. Denote by $s$,
$s=0$ or 1, the number of $x$'s that satisfy  $f(x)=1$. 
The state (\ref{state2}) can be written as
\be
|\psi[2]\rangle
&=&
\frac{1}{\sqrt{2^n}}\sum_{i_2\dots i_n}
|0_1,i_2,\dots,i_n\rangle|f(0_1,i_2,\dots,i_n)\rangle\nonumber\\
&\pp =&+
\frac{1}{\sqrt{2^n}}\sum_{i_2\dots i_n}
|1_1,i_2,\dots,i_n\rangle|f(1_1,i_2,\dots,i_n)\rangle
\ee
Let us note that with very high probability the state is 
\be
\frac{1}{\sqrt{2^n}}\sum_{i_2\dots i_n=0}^1
\Big(|0_1,i_2,\dots,i_n\rangle|0\rangle
+
|1_1,i_2,\dots,i_n\rangle|0\rangle\Big)\label{51}
\ee
With much smaller probability it is either
\be
\frac{1}{\sqrt{2^n}}\sum_{i_2\dots i_n=0}^1
\Big(|0_1,i_2,\dots,i_n\rangle|1\rangle
+
|1_1,i_2,\dots,i_n\rangle|0\rangle\Big)\label{52}
\ee
or
\be
\frac{1}{\sqrt{2^n}}\sum_{i_2\dots i_n=0}^1
\Big(|0_1,i_2,\dots,i_n\rangle|0\rangle
+
|1_1,i_2,\dots,i_n\rangle|1\rangle\Big)\label{53}
\ee
and is never in the form
\be
\frac{1}{\sqrt{2^n}}\sum_{i_2\dots i_n=0}^1
\Big(|0_1,i_2,\dots,i_n\rangle|1\rangle
+
|1_1,i_2,\dots,i_n\rangle|1\rangle\Big)
\ee
since this would mean there are two different numbers satisfying
$f(x)=1$ which contradicts our assumption.
The idea of the algorithm is to apply to the flag qubit 
a nonlinearity that leaves
(\ref{51}) unchanged but (\ref{52}) and (\ref{53}) transforms
into 
\be
\frac{1}{\sqrt{2^n}}\sum_{i_2\dots i_n=0}^1
|0_1,i_2,\dots,i_n\rangle|1\rangle
+
|1_1,i_2,\dots,i_n\rangle|1\rangle\Big).
\label{52'}.
\ee
Although such a dynamics can be approximated by a nonlinear Schr\"odinger
dynamics of a Weinberg type, it is easy to show that it is
unphysical \cite{MCfpl}. 

To see this assume that the flag qubit does not interact with
the $n$ input ones. To simplify the discussion take $n=1$ and
consider the transformation (\ref{mob}). We assume that the
nonlinear evolution is applied {\it locally\/} to the flag
system. (\ref{mob}) implies the following transformation of the
reduced density matrix of the first qubit 
\be
\frac{1}{2}\Big(
|0_1\rangle\langle 0_1|+
|1_1\rangle\langle 1_1|\Big)
\to
\frac{1}{2}\Big(
|0_1\rangle+
|1_1\rangle\Big)
\Big(
\langle 0_1|+
\langle 1_1|\Big)\nonumber
\ee
and therefore a fully mixed state evolves into a pure one. 
It can be shown \cite{MCpla} that Weinberg's description of
separated systems leads to 2-particle Schr\"odinger equations
that induce this kind of behavior at a distance
(``faster-than-light telegraph"). 

Before performing a more detailed analysis 
let us illustrate the crucial element of the algorithm on a simple example.
Take $n=3$ and $f(110)=1$. The oracle produces
\be
{}&{}&8^{-1/2}   \big[ |000\rangle|0\rangle +\nonumber\\
&{}&\pp {8^{-1/2}\big[}|001\rangle|0\rangle +\nonumber\\
&{}&\pp {8^{-1/2}\big[}|010\rangle|0\rangle +\nonumber\\
&{}&\pp {8^{-1/2}\big[}|011\rangle|0\rangle +\nonumber\\
&{}&\pp {8^{-1/2}\big[}|100\rangle|0\rangle +\nonumber\\
&{}&\pp {8^{-1/2}\big[}|101\rangle|0\rangle +\nonumber\\
&{}&\pp {8^{-1/2}\big[}|110\rangle|1\rangle +\nonumber\\
&{}&\pp {8^{-1/2}\big[}|111\rangle|0\rangle \big]\label{ex}
\ee
The nonlinearity now ``looks" at the {\it second\/} and the {\it
third\/} input slots and sees the above kets as the following pairs 
\be
{}&{}&8^{-1/2}   
\big[ |000\rangle|0\rangle +|100\rangle|0\rangle +\nonumber\\
&{}&\pp {8^{-1/2}\big[} 
|001\rangle|0\rangle +|101\rangle|0\rangle +\nonumber\\
&{}&\pp {8^{-1/2}\big[} 
|010\rangle|0\rangle +|110\rangle|1\rangle +\nonumber\\
&{}&\pp {8^{-1/2}\big[} 
|011\rangle|0\rangle +|111\rangle|0\rangle \big]
\ee
Now it scans each of the rows and does not do anything when two
flag 0's occur, but when it ``notices" one 0 and one 1 it changes 0
to 1. So after this step we get 
\be
{}&{}&8^{-1/2}   \big[ 
|000\rangle|0\rangle +|100\rangle|0\rangle +\nonumber\\
&{}&\pp {8^{-1/2}\big[} 
|001\rangle|0\rangle +|101\rangle|0\rangle +\nonumber\\
&{}&\pp {8^{-1/2}\big[} 
|010\rangle|1\rangle +|110\rangle|1\rangle +\nonumber\\
&{}&\pp {8^{-1/2}\big[} 
|011\rangle|0\rangle +|111\rangle|0\rangle \big]
\ee
Now the nonlinearity looks at the {\it first\/} and the {\it
third\/} slots and sees the kets as the following pairs
\be
{}&{}&8^{-1/2}   \big[ 
|000\rangle|0\rangle +|010\rangle|1\rangle +\nonumber\\
&{}&\pp {8^{-1/2}\big[} 
|001\rangle|0\rangle +|011\rangle|0\rangle +\nonumber\\
&{}&\pp {8^{-1/2}\big[} 
|100\rangle|0\rangle +|110\rangle|1\rangle +\nonumber\\
&{}&\pp {8^{-1/2}\big[} 
|101\rangle|0\rangle +|111\rangle|0\rangle \big]
\ee
It again behaves as before and what we get after this step looks
as follows
\be
{}&{}&8^{-1/2} \big[ |000\rangle|1\rangle
+|010\rangle|1\rangle +\nonumber\\ 
&{}&\pp {8^{-1/2}\big[} |001\rangle|0\rangle
+|011\rangle|0\rangle +\nonumber\\ 
&{}&\pp {8^{-1/2}\big[} |100\rangle|1\rangle
+|110\rangle|1\rangle +\nonumber\\ 
&{}&\pp {8^{-1/2}\big[} |101\rangle|0\rangle
+|111\rangle|0\rangle \big] 
\ee
Finally our nonlinearity looks at the {\it first\/} and the {\it
second\/} slots and the state regroups in the following way
\be
{}&{}&8^{-1/2} \big[ |000\rangle|1\rangle
+|001\rangle|0\rangle +\nonumber\\ 
&{}&\pp {8^{-1/2}\big[} |010\rangle|1\rangle
+|011\rangle|0\rangle +\nonumber\\ 
&{}&\pp {8^{-1/2}\big[} |100\rangle|1\rangle
+|101\rangle|0\rangle +\nonumber\\ 
&{}&\pp {8^{-1/2}\big[} |110\rangle|1\rangle
+|111\rangle|0\rangle \big] 
\ee
Now each row contains one 1 and in the final move all flag 0's
are switched to 1's and the state partly disentangles:
\be
{}&{}&8^{-1/2}   \big[ |000\rangle +|001\rangle +\nonumber\\
&{}&\pp {8^{-1/2}\big[} |010\rangle +|011\rangle +\nonumber\\
&{}&\pp {8^{-1/2}\big[} |100\rangle +|101\rangle +\nonumber\\
&{}&\pp {8^{-1/2}\big[} |110\rangle +|111\rangle \big]|1\rangle
\ee
Of course, in case $s=0$ the entire state does not change during
the operation and a measurement on the flag qubit gives 0 with
certainty. Such an algorithm is fast and allows to distinguish
between $s=0$ and $s\neq 0$ in a linear time. 
The number of operations is of the
order of $n$ as compared to $2^n$ typical of a slow
data-search algorithm. 

It follows that we have an algorithm that is fast but
simultaneously makes an explicit use of an arbitrarily fast,
physically unacaptable process. 
Now I will show that the faster-than-light effect can be
eliminated without any loss in the efficiency of the algorithm. 

To do so I will use an explicit nonlinear dynamics and apply it
{\it locally\/} to the flag qubit. By saying that the dynamics
is applied locally it is meant that we are using an appropriate
$(n+1)$-particle extension of a nonlinear 1-particle dynamics.
Assume the $n+1$ subsystems do not interact with one another.
The extension is local if it satisfies the following condition:
A reduced density matrix of any of the $n+1$ subsystems
satisfies a Liouville-von Neumann (nonlinear) equation which
contains the reduced density matrix and Hamiltonian of {\it only
this\/} particular subsystem. In the case of Weinberg's
nonlinear quantum mechanics of pure states and for
finite-dimensional systems 
the extension of this type was introduced by Polchinski
\cite{Polchinski}. Its generalization to mixed states was given
in \cite{Bona,Jordan}. The extension to more
general theories was discussed in \cite{MCqph,MCijtp} and applied to
concrete problems in \cite{MCpra,MCMK}. The algebraic origin of
the Polchinski-B\'ona-Jordan formulation was discussed in detail
in \cite{MCpla} and \cite{MCijtp}. It should be noted that
extensions of this type are nonunique if one starts with a
nonlinear dynamics of state vectors (cf.
\cite{Polchinski,MCpra}). They become unique if the dynamics is
from the outset given in terms of density matrices. From the
point of view of the algorithm the uniqueness problem is
irrelevant so we can stick to the nonlinear Schr\"odinger
equation framework. 

\section{Modification of the fourth step}

The above procedure can be, in principle, implemented in terms
of a very complicated Schr\"odinger-type dynamics of the
entire $(n+1)$-particle system of the quantum computer. It 
cannot be achieved by applying the nonlinear evolution locally
to the flag qubit without generating the unphysical influences
between different parts of the computer.
 
Below I propose a simpler procedure which is based on a
nonlinearity which is applied only to the flag qubit.

Let us begin with a 1-qubit system whose dynamics is
described by the nonlinear Schr\"odinger equation
\be
i|\dot \psi\rangle &=&
\epsilon\tanh\Big(\alpha\langle\psi|A-\eta\bbox 1
|\psi\rangle\Big)A |\psi\rangle
\ee
where $\epsilon$ is the magnitude of the nonlinearity, $\alpha$
is a very large real number (say, $\alpha\approx 2^n$),
\be
A=\eta \Big(|0\rangle\langle 0|
- |1\rangle\langle 1|\Big)
+
\sqrt{1-\eta^2}\Big(|0\rangle\langle 1|+|1\rangle\langle
0|\Big),
\ee
and $\eta$ is small but nonzero.
For $|\psi\rangle=|0\rangle$ the expression
under $\tanh$ vanishes. For a small admixture of
$|1\rangle$ and sufficiently large $\alpha$ the mobility
with a nonzero frequency begins and an arbitrarily small amount of
$|1\rangle$ can be sufficiently amplified. The Polchinski-type 
local extension of the dynamics
to the entire quantum computer is \cite{extension}
\be
i|\dot \Psi\rangle &=&
\epsilon\tanh\Big(\alpha\langle\Psi|\bbox 1^{(n)}\otimes(A-\eta\bbox
1)|\Psi\rangle\Big)\bbox 1^{(n)}\otimes A |\Psi\rangle.\nonumber
\ee
The $(n+1)$-particle solution is
\be
|\Psi_t\rangle &=&
\Big(
\bbox 1^{(n+1)} \cos  \omega t -i
\bbox 1^{(n)}\otimes A\sin
 \omega t \Big)|\Psi_0\rangle 
\ee
with 
\be
\omega
&=&
\epsilon\tanh\Big(\alpha{\rm Tr}\rho(A-\eta\bbox
1)\Big)\nonumber\\
&=&\epsilon\tanh\Big(\frac{\alpha\eta
s}{2^{n-1}}\Big)
\ee
where 
\be
\rho={\rm Tr}_{1\dots n}|\Psi_0\rangle \langle\Psi_0|
=
\frac{2^n-s}{2^n}
|0\rangle\langle 0|+
\frac{s}{2^n}
|1\rangle\langle 1|\nonumber
\ee
is the reduced density matrix of the flag system after the first
three steps of the original Abrams-Lloyd algorithm. 

The average of
$\sigma_3=|0\rangle\langle 0|- |1\rangle\langle 1|$ at the flag
subsystem is 
\be
\langle\sigma_3\rangle &=&\langle\Psi_t|\bbox
1^{(n)}\otimes\sigma_3|\Psi_t\rangle
\nonumber\\
&=&
\frac{2^{n-1}-s}{2^{n-1}}\cos 2 \omega t 
+
2\eta^2\frac{2^{n-1}-s}{2^{n-1}}\sin^2 \omega t
\ee
For $s=0$ the average is constant in time and equals 1. For
$s=1$, $\eta^2\approx 0$, and sufficiently large $\alpha$
it oscillates with $\omega\approx\epsilon$.
For $t\approx \pi/\epsilon$ the average is
$\langle\sigma_3\rangle \approx -1$, which means that almost all
flag 0's in (\ref{state2}) have been changed to 1's.
Therefore instead of applying a complicated nonlinear dynamics
of the original ``Step~4" one can use the fact that in the {\it
local\/} description the 1-particle nonlinearity is sensitive to
the {\it reduced density matrix\/} of the particle.

This kind of algorithm cannot distinguish between different
nonzero values of $s$, but clearly distinguishes between $s=0$
and $s\neq 0$ in a way that is insensitive to small fluctuations of
the parameters. 
It is interesting that the modified ``Step~4" is essentially
non-algorithmic and more resembles an effect of
puncturing a baloon than performing an algorithm. 

\acknowledgements

I am indebted to J.~Naudts and S.~Lloyd for 
comments, M.~Kuna for pointing out some errors, A.~Posiewnik for
his remarks on quantum computation vs. mobility, and my students
from the Computer Physics Section of Politechnika Gda\'nska for
a stimulating semester of discussions on quantum computers. 
My work was supported by the Polish-Flemish grant 007 and was
done mainly during my stay at the University of Antwerp.


\begin{references}
\bibitem{extension}The extension proposed in \cite{Polchinski,Bona,Jordan}
referred to those nonlinear Schr\"odinger equations that possess
Hamiltonian functions. There exist nonlinear equations that do
not have this property (e.g non-linearizable Doebner-Goldin
equations \cite{DG}) and therefore the standard results cannot
be directly applied to them. The discussion given in
\cite{MCqph} referred to such more general situations. It was
shown in \cite{MCqph} that the local extension involves, in
general, integro-differential equations which have been
typically discarded as nonlocal. Such nonlocally looking
equations generate, paradoxically, local physics.
\end{references}
\end{document}